# Inducing electron spin coherence in GaAs quantum well waveguides: Spin coherence without spin precession

Susanta Sarkar, Phedon Palinginis, and Hailin Wang

Department of Physics, University of Oregon, Eugene, OR 97403, USA

Pei-Cheng Ku and Connie J. Chang-Hasnain

Department of Electrical Engineering and Computer Science

University of California, Berkeley, CA 94720, USA

N.H. Kwong and R. Binder

Optical Sciences Center, University of Arizona, Tucson, AZ 85721, USA

Abstract

Electron spin coherence is induced via light-hole transitions in a quantum well waveguide without either an external or internal DC magnetic field. In the absence of spin precession, the induced spin coherence is detected through effects of quantum interference in the spectral domain coherent nonlinear optical response. We interpret the experimental results qualitatively using a simple few-level model with only the optical transition selection rule as its basic ingredients.

Coherent manipulation of electron spins in semiconductors plays a central role in spin-based electronics and photonics and in spin-based quantum information technologies [1]. Extensive experimental studies in this area have led to the demonstration of remarkable phenomena such as the persistence of robust electron spin coherence up to room temperature and the transfer of electron spin coherence between molecularly bridged quantum dots [2, 3]. Most of the experimental efforts have thus far been carried out in the presence of an external DC magnetic field in a Voigt geometry. The external magnetic field induces a Zeeman splitting between the two conduction band electron spin states. Coherent superposition of these spin states, i.e. electron spin coherence, corresponds to a Larmor precession of the electron spin and can be probed with transient optical techniques. Coherent spin manipulation without external magnetic fields in non-magnetic semiconductors has also been demonstrated in a recent study [4], in which an effective internal magnetic field induced by strain leads to a spin splitting in the conduction band.

In this paper, we report the experimental demonstration and simple theoretical analysis of inducing and detecting electron spin coherence in a GaAs quantum well (QW) without the use of either an *external* or *internal* magnetic field. We have taken advantage of the spin-orbit coupling in the valence band and have used light-hole (LH) transitions in a waveguide geometry to induce a coherent superposition of electron spin states. The induced spin coherence exhibits no spin precession. We have used quantum interference in spectral domain, instead of time domain, coherent optical response to probe the spin coherence.

The band edge in semiconductors such as GaAs is characterized by a *s*-like conduction band and a *p*-like valence band. Spin-orbit coupling in the valence band leads to the formation of heavy-hole (HH) and LH valence bands with $j_z= \pm 3/2$ and $\pm 1/2$, respectively. Quantum confinement in a QW further lifts the degeneracy between the HH and LH valence bands at the band edge. Figure 1a shows the optical selection rule for dipole transitions between the conduction and the HH and LH valence bands in a GaAs QW.

For the LH valence bands, the spin-orbit coupling mixes spin-up and spin-down states, which makes it possible to couple the $s_z = \pm 1/2$ conduction bands to a common LH valence band via two dipole optical transitions (see Fig. 1a). We can induce a coherent superposition of the electron spin states by exciting both of these two transitions. As indicated in Fig. 1a, one of the transitions is for optical fields polarized along the $z$-direction [5]. For a (001) QW, the $z$-polarized field has to propagate in the plane of the QW. The optical excitation thus has to be carried out in a waveguide geometry.

In comparison, in a normal geometry, optical excitations of electron spin coherence through either HH or LH transitions are possible only in the presence of an external or an effective internal DC magnetic field in the plane of the QW (Voigt geometry). As discussed earlier, the electron spin coherence in this case corresponds to a Larmor spin precession around the DC magnetic field with the precession frequency determined by the electron spin splitting. Coherent oscillations arising from the spin precession can be detected with transient optical techniques, such as time-resolved Faraday rotation, time-resolved photoluminescence, or transient differential absorption [6-8].

In the absence of external or internal DC magnetic fields, no spin precession is present. Transient optical techniques are no longer effective for probing the electron spin coherence. Electron spin coherence, however, can still lead to quantum interference, inducing a transmission resonance in the spectral domain coherent optical response, as we will discuss in detail later. The linewidth of the induced resonance corresponds to the decoherence rate for the electron spin coherence.

Experimental studies were carried out in a slab waveguide consisting of a 17.5 nm GaAs/AlGaAs (001) QW and GaAs/AlGaAs superlattice cladding layers grown by molecular beam epitaxy. The waveguide was cleaved to a length of nearly 100 μm and was mounted on a bridge-like brass sample holder attached to the cold finger of an optical cryostat. A 20X optical objective (Mitutoyo) was used to couple the incident laser beams into the waveguide. An aspheric lens (numerical aperture=0.44) mounted on a miniature translation stage was also placed inside the cryostat for collimation of output from the back facet of the waveguide. Figures 1b and 1c show the linear transmission spectra of the waveguide for TE (electric field parallel to the QW plane) and TM (electric field perpendicular to the QW plane) polarized optical fields, respectively. As expected from the polarization selection rule, for the TE polarization both HH and LH exciton resonances are observed, whereas for the TM polarization HH exciton resonance is nearly negligible [5]. Fabry-Perot fringes due to reflection between the two facets of the waveguide are also clearly visible in the transmission spectra. Note that strong exciton absorption ($\alpha L$>10) leads to transmission resonances with a nearly flat bottom. To perform differential transmission (DT) measurements, we used a tunable diode laser (from New Focus) and a tunable Ti:Sapphire laser (Coherent 899) as the probe and pump, respectively. The relative frequency jitter between the pump and probe is less than 20 MHz. Both beams propagate along the same direction in the waveguide. To avoid strong excitonic optical absorption in the waveguide, the pump beam was fixed at a spectral position 0.5 to 1 nm below the LH exciton absorption line center. To single out the DT response, we used lock-in detection with dual modulation. The intensity of the pump beam was modulated at frequency $\omega_1$ and that of the probe was modulated at frequency $\omega_2$. The DT response, with an intensity modulation frequency of $|\omega_2 - \omega_1|$, was then measured with a lock-in amplifier.

The basic idea of the experiment can be understood with the help of Fig. 1a. For the excitation of the electron spin coherence, the pump with TE-polarization and the probe with TM-polarization couple the two electron spin states in the conduction band to a common LH valence band state (TE-polarized fields consist of both σ+ and σ- circularly-polarization fields in Fig. 1a). The spin coherence is formed to the second order of the applied optical field (linear to both the pump and probe field). The pump further interacts with the induced spin coherence, generating a third order nonlinear optical response. This spin-coherence-induced nonlinearity can be detected in the DT response, i.e, the change in the probe transmission induced by the pump. Specifically, we will show below that the DT response as a function of the pump-probe detuning exhibits an induced resonance associated with the electron spin coherence.

Figure 2a shows the DT response obtained at T=50 K. A sharp induced resonance occurs at the zero pump-probe detuning. The spectral linewidth of the resonance (FWHM) is 1 GHz, corresponding to a decay time of 300 ps, in agreement with an earlier transient measurement of the spin decoherence time in a similar 17.5 nm GaAs QW [8]. Figure 2b also shows the dependence of the amplitude of the induced resonance on the input pump intensity. At the intensity range used for the measurement, the amplitude scales linearly with the pump intensity.

To confirm that the induced resonance in Fig. 2a indeed arises from the electron spin coherence, we carried out additional experimental studies, in which we applied to the QW waveguide an external magnetic field (B=0.25 T) along the z-axis. In this Faraday geometry, the external magnetic field induces energy splitting in both the conduction and the LH valence bands, but does not affect the dipole optical selection rule. The electron spin coherence is thus excited through the same mechanism regardless whether there is an external magnetic field (this would not be the case if the external magnetic field were applied in the Voigt geometry). We stress that while the use of the external magnetic field enables us to demonstrate and clarify the physical origin of the induced resonance in the DT response, the excitation and detection of the electron spin coherence do not rely on the presence of the external magnetic field.

Figures 3a and 3b show the DT responses obtained at T=50 K and T=20 K, respectively. Two induced resonances now occur symmetrically away from the zero pump-probe detuning. The spectral position of the induced resonances corresponds to the electron Zeeman splitting (the electron g-factor was determined in an earlier transient measurement [8]), demonstrating that the induced resonance arises directly from the electron spin coherence. In the temperature range used for Figs. 3a and 3b, the linewidth of the induced resonance and thus the corresponding spin decoherence rate actually decreases with increasing temperature, as expected from earlier experimental studies [9]. Note that, in principle, coherent superposition of the two LH valence bands can also contribute to the overall DT response. Contributions from the inter-valence band coherence, however, are negligible due to the rapid spin decoherence associated with the holes.

For comparison, we show in Fig. 3c the DT response obtained with the pump and probe having the same TE-polarization and with otherwise identical conditions to Fig. 3b (similar results were also obtained when both the pump and probe are TM-polarized). As expected, the spin-coherence induced resonances vanish since in this case the pump and probe cannot couple the two electron spin states to a common LH valence band. Instead, a sharp resonance is observed at zero pump-probe detuning. This induced resonance is due to the well-known phenomenon of exciton population oscillation and the linewidth of the resonance is determined by the exciton lifetime [10]. Note that the dipole optical selection rule in Fig. 1a dictates that exciton population oscillation cannot occur when the pump and probe are TE and TM polarized, respectively, which is shown in the following theoretical analysis and is also confirmed experimentally by the absence of an induced resonance at zero pump-probe detuning in Figs. 3a and 3b.

As discussed above, the sharp resonances in the DT response induced by the electron spin coherence is a direct consequence of the optical transition selection rules of our system. To demonstrate this, we used a simple few-level model for the QW system. This model consists of the two conduction band states and the two LH valence band states in Fig. 1a. The model Hamiltonian assigns an energy to each of the four states and optical transitions between the states as shown in Fig. 1a. A weak magnetic field in the Faraday geometry, when switched on, leads to a Zeeman-splitting of the conduction band denoted by $2\hbar\Delta_e$. The splitting in the LH band, is denoted by $2\hbar\Delta_{lh}$. We solve the density-matrix equations of motion under this Hamiltonian at steady state and with the two valence band states initially filled and the two conduction band states initially empty. The DT response to the second order of the pump field and first order of the probe field (see e.g. [11]) and with the pump TM-polarized and the probe TE-polarized is given by

$$\Delta T^{(3)} \propto -\frac{|\Omega_t|^2 |\Omega_p|^2}{8} \text{Im}(A_+ + A_-) \quad , \tag{1}$$

$$A_\pm = \frac{1}{\delta_t \mp (\Delta_e + \Delta_{lh}) + i\gamma} \left[ \frac{1}{\delta_t - \delta_p \mp 2\Delta_e + i\gamma_s} \left( \frac{1}{\delta_p \pm (\Delta_e - \Delta_{lh}) - i\gamma} - \frac{1}{\delta_t \mp (\Delta_e + \Delta_{lh}) + i\gamma} \right) + \frac{4\gamma}{\Gamma} I_\pm \right]$$

Here $\delta_p$ ($\delta_t$) is the detuning of the pump (probe) frequency from the LH transition frequency at zero magnetic field, $\gamma$ is the optical transition linewidth, $\gamma_s$ is the decay rate of the electron spin coherence, $\Gamma$ is the life time of the conduction band levels, and $I_\pm = 1/\{\gamma^2 + [\delta_p \mp (\Delta_e - \Delta_{lh})]^2\}$. The Rabi frequencies for both optical fields are defined with the dipole moment for circular polarizations: $|\hbar\Omega_{p/t}| = |d_+||\vec{E}_{p/t}|$.

There are two types of contributions to the DT response (the contribution of the short-lived coherence between the two LH states is ignored in Eq. 1). The first type is represented by the last term in $A_\pm$ and is due to incoherent bleaching. This contribution leads to a spectrally broad response. The second type, represented by the first term in $A_\pm$, is due to the contribution of the electron spin coherence. With the pump TM-polarized and the probe TE-polarized, our model is a double-V system with the two conduction band spin states as the two excited states in each V-type 3-level system. The optical transition path that excites the spin coherence interferes destructively with the direct (first order) probe transition in each V-system. Similar to electromagnetically induced transparency, this destructive interference results in a sharp induced transmission resonance in the DT response [12]. As shown in Eq. 1, the spectral position of the induced resonance in the DT response is determined by the two-photon resonance condition $\delta_p - \delta_t \pm 2\Delta_e = 0$ and the width of the resonance is determined by the decay rate of the spin coherence. In the absence of an external magnetic field, the induced resonance occurs at zero pump-probe detuning.

When both the pump and probe are TE-polarized, our model becomes that of two decoupled two-level systems. The DT response is then given by

$$\Delta T^{(3)} \propto -\frac{|\Omega_t|^2 |\Omega_p|^2}{16} \text{Im}(B_+ + B_-), \tag{2}$$

$$B_{\pm} = \frac{1}{\delta_t \mp (\Delta_e - \Delta_{lh}) + i\gamma} \left[ \frac{1}{\delta_t - \delta_p + i\Gamma} \left( \frac{1}{\delta_p \mp (\Delta_e - \Delta_{lh}) - i\gamma} - \frac{1}{\delta_t \mp (\Delta_e - \Delta_{lh}) + i\gamma} \right) + \frac{2\gamma}{\Gamma} I_{\pm} \right]$$

In this configuration, $\Delta T^{(3)}$ consists of a spectrally broad response due to incoherent bleaching and also a sharp resonance at zero pump-probe detuning. No spin coherence can be created in this case; instead, the sharp resonance is set up by the two optical fields coupling to the same transition: it is the result of a coherent population oscillation [11]. As such, the width of the resonance is determined by the excited state lifetime $1/\Gamma$. Note that the sharp resonance occurs at zero pump-probe detuning even in the presence of an external magnetic field.

In Fig. 4, we show numerical results from Eqs. 1 and 2 with parameters given by the experiment: $\gamma = .16$ THz, $\gamma_s = \Gamma = 1$ GHz, $\Delta_e = .75$ GHz, $\delta_p = 48$ GHz. For convenience, we also choose $\Delta_{lh} = 0$. Our simple model describes qualitatively the principal resonance features of the experiment, confirming that the sharp resonance in the DT response for TE pump and TM probe is induced by an electron spin resonance. While the simple theory based on selection rules yields good qualitative agreement with the experiment, an improved quantitative agreement should be possible with fully microscopic theories, for example along the lines of Ref. [13].

In summary, by taking advantage of spin-orbit coupling in the valence band and by using light-hole transitions in a waveguide geometry, we have demonstrated the excitation of electron spin coherence without either an external or internal DC magnetic field. We show that in the absence of spin precession, the electron spin coherence can be detected through effects of quantum interference in spectral domain coherent nonlinear optical response. The generation and detection of electron spin coherence without the complexity and/or added decoherence of external or effective internal magnetic fields can potentially open up new avenues for spintronics and spin-based photonics, namely the creation, manipulation and utilization of spins in semiconductors.

This work is supported by NSF-DMR (HW), ARO, DARPA-SPINS/ONR, DARPA university photonics research center, JSOP, and AFOSR.

Figure captions:

FIG. 1 (a) Optical selection rule for dipole transitions between the conduction and the HH and LH valence bands in a GaAs QW. (b) and (c) Linear transmission spectra of the QW waveguide for TE and TM polarized fields.

FIG. 2 (a) Differential transmission response obtained with $I_{pump}$= 2 mW and $I_{probe}$= 0.25 mW. The pump and probe are TE and TM polarized, respectively. The solid line is a fit to Lorentzian. (b) The amplitude of the induced resonance as a function of the input pump power with $I_{probe}$=0.25 mW. The solid line is a guide to the eye.

FIG. 3 Differential transmission response obtained with $I_{pump}$= 2 mW and $I_{probe}$= 0.25 mW and with an external magnetic field (B=0.25 T) applied in the Faraday geometry. For (a) and (b), the pump and probe are TE and TM polarized, respectively. For (c), both the pump and probe are TE polarized.

FIG. 4 Third order differential transmission response calculated within the model described in the text. (a) Pump is TE polarized and probe is TM polarized. (b) Both pump and probe are TE polarized. See text for parameter values used.

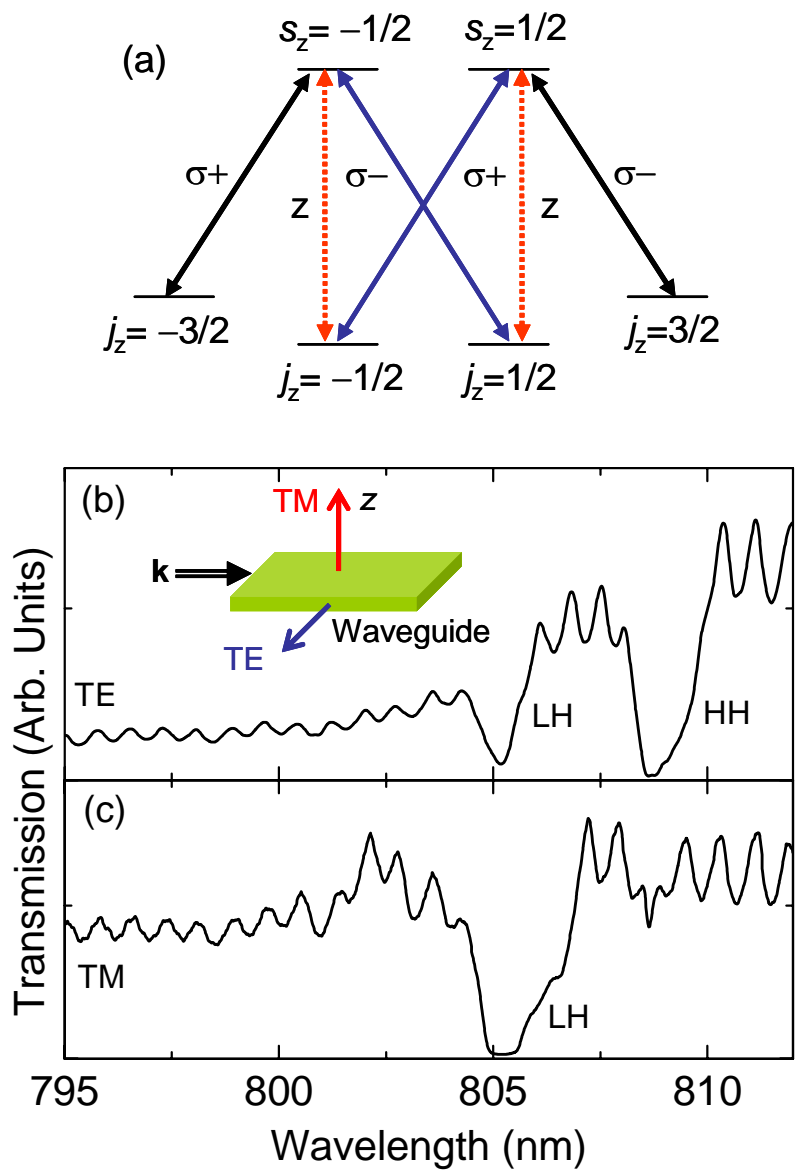

Fig. 1

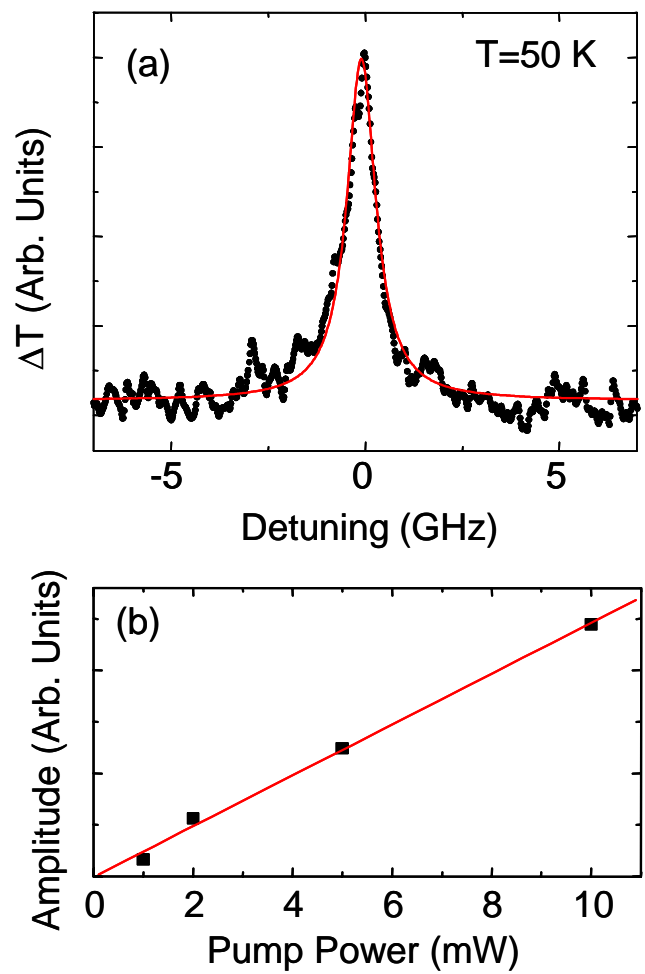

Fig. 2

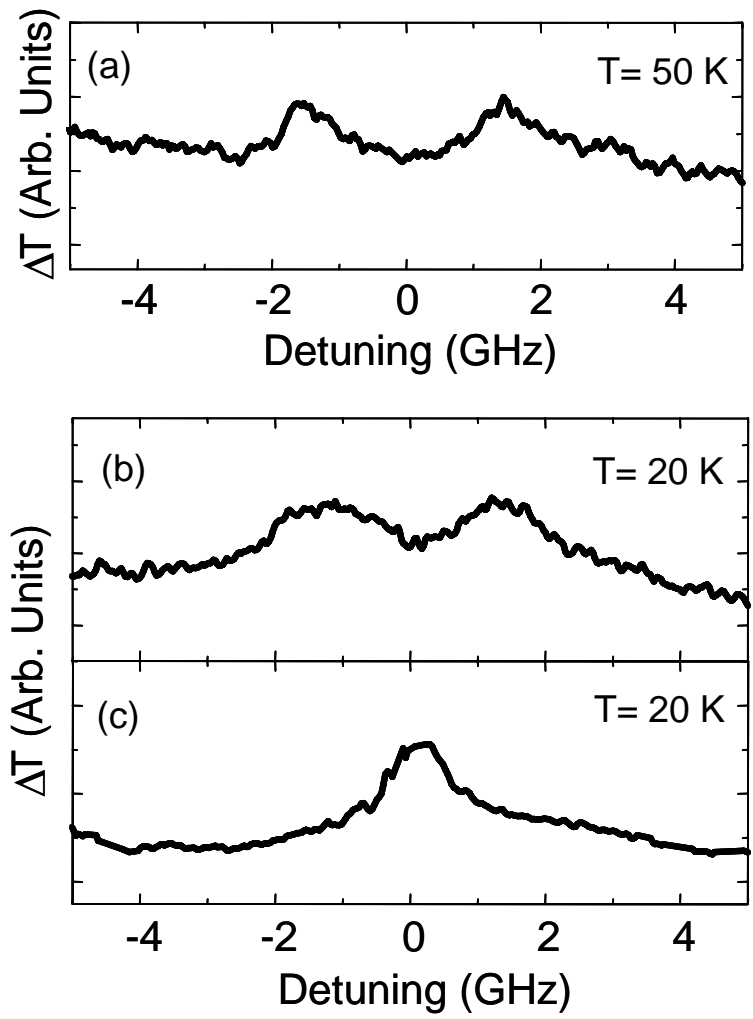

Fig. 3

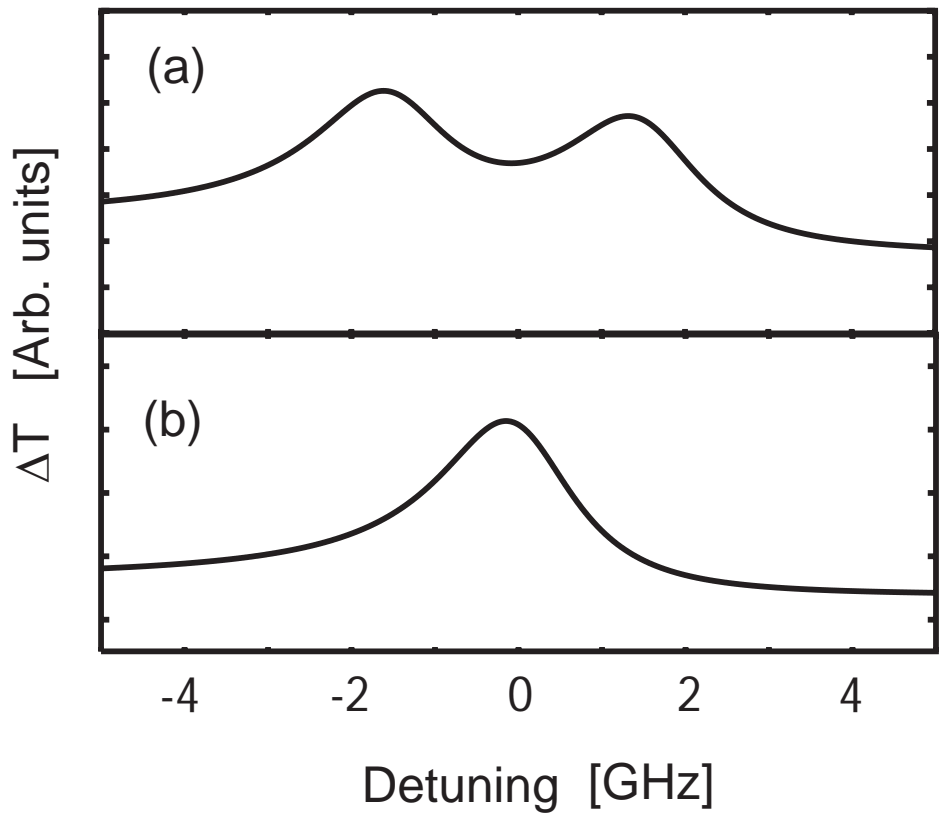

Fig. 4